\documentclass[a4paper,11pt]{article}
\pdfoutput=1
\usepackage{amsmath}
\usepackage{amssymb}
\usepackage{color}
\usepackage{cite}
\usepackage{graphicx}
\usepackage{leftidx}
\usepackage{subfig}
\usepackage{hyperref}

\numberwithin{equation}{section}

\def\beq{\begin{equation}}
\def\eeq{\end{equation}}

\def\l{\lambda}

\def\o{\omega}
\def\O{\Omega}

\def\ra{\rangle}
\def\d{\mathrm{d}}

\newcommand{\ben}{\begin{enumerate}}
\newcommand{\een}{\end{enumerate}}
\newcommand{\be}{\begin{equation}}
\newcommand{\ee}{\end{equation}}
\newcommand{\mc}{\mathcal}

\newcommand{\de}{\delta}
\newcommand{\del}{\partial}
\newcommand{\w}{\wedge}
\newcommand{\da}{\dagger}

\usepackage{tikz}
\usetikzlibrary{positioning}
\usetikzlibrary{intersections}
\usetikzlibrary{fadings} 
\usetikzlibrary{arrows.meta} 
\usetikzlibrary{arrows}

\tikzfading[name=fade out,
inner color=transparent!0,
outer color=transparent!100]

\definecolor{cherryblossompink}{rgb}{1.0, 0.72, 0.77}
\definecolor{lightblue}{rgb}{0.68, 0.85, 0.9}

\usetikzlibrary{decorations.pathmorphing}
\usetikzlibrary{decorations.pathreplacing,decorations.markings}

\usetikzlibrary{backgrounds,automata}

\begin{document}

\title{\textbf{Canonical quantization of the Pais-Uhlenbeck oscillator with a higher-derivative perturbation: a covariant phase space approach}}
\author{
Jie-qiang Wu$^{1,2}$\footnote{jieqiangwu@itp.ac.cn}\, and
Jinan Zhao$^{1,2}$\footnote{jinanzhao@itp.ac.cn}
}
\date{\today}

\maketitle

\begin{center}
{\it
$^{1}$Institute of Theoretical Physics, Chinese Academy of Sciences, Beijing 100190, China \\
$^{2}$School of Physical Sciences, University of Chinese Academy of Sciences, \\ Beijing 100049, China
}
\vspace{10mm}
\end{center}

\begin{abstract}

In this paper, we apply the covariant phase space formalism to the perturbative canonical quantization of the Pais-Uhlenbeck oscillator, with the acceleration-squared term treated as a perturbation. We quantize this model by constructing the symplectic form on the low-energy solution space. We then compute the energy spectrum and the unequal-time commutator in a perturbative way, and obtain the results that agree with the expansion of the exact low-energy theory. The perturbation method bypasses the standard Ostrogradsky construction and naturally decouples the Ostrogradsky ghost. This work extends our previous perturbative quantization scheme to genuine higher-derivative theories.

\end{abstract}

\baselineskip 18pt
\thispagestyle{empty}

\newpage

\tableofcontents

\section{Introduction}\label{Sec1}

Higher-derivative theories and nonlocal theories have long been a subject of interest in theoretical physics. They were originally introduced in attempts to construct finite quantum field theories~\cite{Pais:1950za}, arise naturally in string theory~\cite{Eliezer:1989cr,Moeller:2002vx}, and find applications in higher-derivative gravity~\cite{Stelle:1976gc,Julve:1978xn,Nojiri:2001ae,Woodard:2006nt,Langlois:2015cwa,Lu:2015cqa,Shankaranarayanan:2022wbx}, meson-nucleon interactions~\cite{KristensenMoller1952}, non-commutative field theories~\cite{Connes:1997cr,Seiberg:1999vs,Douglas:2001ba,Szabo:2001kg}, and many other contexts.

Ostrogradsky's construction provides the standard Hamiltonian formalism for higher-derivative theories~\cite{Ostrogradsky1850}. A classic result of this construction is the Ostrogradsky theorem: 
for any non-degenerate Lagrangian containing higher-order time derivatives, the Hamiltonian is necessarily linear in some of the canonical momenta, and is therefore unbounded from below~\cite{Woodard:2015zca}. This instability, known as the Ostrogradsky ghost, renders the classical theory unstable and leads to vacuum decay in the quantum theory. Consequently, in the effective field theory framework, higher-derivative terms are usually treated as perturbations, 
and ghost modes are rendered invisible in the low-energy regimes.

Various schemes have been proposed to achieve a perturbative treatment of higher-derivative theories.
These include field redefinitions~\cite{Barua:1977is}, constraint-based reductions~\cite{Jaen:1986iz,Simon:1990ic,Reyes:2009zb,Chen:2012au}, and order-by-order constructions of an effective canonical structure~\cite{Cheng:2001du,Cheng:2002rz}. A common theme of most perturbative approaches that the canonical structure should be restricted to the low-energy sector of the phase space.

In a recent work~\cite{Wu:2026xgd}, we demonstrated that the covariant phase space (CPS) formalism~\cite{Lee:1990nz,Wald:1993nt,Iyer:1994ys,Iyer:1995kg,Wald:1999wa,Harlow:2019yfa} provides a systematic framework for the perturbative canonical quantization of systems in which a velocity-squared perturbation alters the constraint structure. The present paper extends this approach to genuine higher-derivative perturbations. We illustrate the method using the Pais-Uhlenbeck (PU) oscillator\cite{Pais:1950za}, with its higher-derivative term viewed as a perturbation. We show that the perturbative CPS quantization naturally selects the physical low-frequency branch and automatically decouples the ghost. We obtain the energy spectrum and the unequal-time commutator that agree with the expansion of the exact solution, which demonstrates that the CPS method offers a viable quantization scheme for higher-derivative theories.

The rest of this paper is organized as follows: In Sec.~\ref{Sec2} we present the Lagrangian of this model. In Sec.~\ref{Sec3} we review the standard Ostrogradsky quantization. In Sec.~\ref{Sec4} we reformulate the quantization with CPS and treat the higher-derivative term as a perturbation. We summarize our results and discuss their implications in Sec.~\ref{Sec5}. In App.~\ref{app:equivalence} we establish the equivalence between the CPS and the Ostrogradsky's construction. Calculation details omitted in the main text are collected in Appendices~\ref{app: SF} and \ref{app: H}.

In this paper, we use natural units with $\hbar=1$. We add dots above functions to represent time derivatives, i.e., $\dot{x}(t) = \frac{\d x(t)}{\d t}$, $\ddot{x}(t) = \frac{\d^2 x(t)}{\d t^2}$. We denote the exterior derivative on phase space by $\de$ to distinguish it from the exterior derivative $\d$ on spacetime.


\section{Lagrangian}\label{Sec2}

The Lagrangian of this model reads\footnote{This Lagrangian is equivalent to the standard Pais-Uhlenbeck form up to a sign convention and parameter redefinition.}
\begin{equation}\label{Lag}
    \mathcal{L}(x,\dot{x}, \ddot{x}) = 
     \frac{1}{2} \dot{x}^2
     - \frac{\o^2}{2}  x^2
     - \frac{\l}{2} \ddot{x}^2,
\end{equation}
where $\o> 0$ and $\l \ge 0$. 
In the canonical quantization procedure of this system, we would like to treat the $\l$-term in the above Lagrangian as a perturbation, and work in the weak-coupling limit $\l \o^2 \ll 1$.


\section{Standard canonical quantization}\label{Sec3}

In this section we review the standard Ostrogradsky's construction for this model~\cite{Woodard:2015zca,Ketov:2011re}. This serves two purposes. First, it demonstrates explicitly how the Ostrogradsky ghost arises in the conventional canonical approach, rendering the Hamiltonian unbounded below. Second, the expansion of the exact low-frequency branch in the limit $\l \to 0$ provides a benchmark against which the perturbative results in Sec.~\ref{Sec4} should be tested.

\subsection{Ostrogradsky's construction}

The equation of motion (EOM) of this model is a fourth-order equation. Thus solutions depend on four pieces of initial value data. There must be four canonical coordinates. Ostrogradsky's choices for these are
\begin{eqnarray}
X_1 &\equiv x , & 
\qquad 
P_1 \equiv \frac{\del \mc{L}}{\del \dot{x}}
- \frac{\d}{\d t} \frac{\del \mc{L}}{\del \ddot{x}}
= \dot{x} + \l \dddot{x} , \label{ct1} \\
X_2&  \equiv \dot{x}, & \qquad 
P_2 \equiv \frac{\del \mc{L}}{\del \ddot{x}}
= - \l \ddot{x}. \label{ct2}
\end{eqnarray}
The Ostrogradsky Hamiltonian is 
\begin{equation}
    H(X_1,X_2,P_1,P_2) \equiv 
    P_1 \dot{X}_1 + P_2 \dot{X}_2 - \mc{L} 
    = P_1 X_2 - \frac{1}{2 \l } P_2 ^2 - \frac{1}{2} X_2^2
    + \frac{1}{2} \o^2 X_1^2.
\end{equation}

\subsection{Quantization}

The EOM of this model is
\begin{equation}\label{EOM}
    -\l \ddddot{x} - \ddot{x} - \o^2 x = 0.
\end{equation}
The solution to EOM can be expanded as 
\begin{equation}
    x(t) = \sqrt{\frac{1}{2 f_+ \sqrt{1 - 4 \l \o^2}}}
    (a_+ e^{-i f_+ t} + a_+^\da e^{i f_+ t})
    + \sqrt{\frac{1}{2 f_- \sqrt{1 - 4 \l \o^2}}}
    (a_-^\da e^{-i f_- t} + a_- e^{i f_- t}),
\end{equation}
where two characteristic frequencies are
\begin{equation}\label{fre}
    f_{\pm} = 
    \sqrt{ \frac{1 \mp \sqrt{1 \!-\! 4 \l \o^2}}{2 \l } }.
\end{equation}
To quantize the model, we impose the following equal-time commutation relations
\begin{align}
    [X_i, P_j] &= i \de_{ij}, \\
    [X_i, X_j] &= [P_i, P_j] = 0.
\end{align}
From which we obtain commutation relations between operators $a_+$, $a_+^\da$, $a_-$ and $a_-^\da$
\begin{align}
    [a_+, a_+^\da] &= [a_-, a_-^\da] = 1, \\
    [a_+, a_-^\da] = [a_+, a_-^\da] 
    &= [a_+^\da, a_-] = [a_+^\da, a_-^\da] = 0.
\end{align}
After quantization, the Hamiltonian operator can be re-expressed in terms of operators $a_+$, $a_+^\da$, $a_-$ and $a_-^\da$ as
\begin{equation}\label{full_Ham}
    H = f_+( a_+^\da a_+ + \tfrac{1}{2}) 
    - f_- ( a_-^\da a_- + \tfrac{1}{2}).
\end{equation}
We define number operators $N_+ = a_+^\da a_+$ and $N_- = a_-^\da a_-$, and label their common eigen-states by $| n_+, n_- \ra$, where $n_+$ and $n_-$ are non-negative integers. Then the energy spectrum is unbounded below
\begin{equation}
    H | n_+, n_- \ra 
    = \big[ f_+( n_+ + \tfrac{1}{2}) 
    - f_- ( n_- + \tfrac{1}{2}) \big] | n_+, n_- \ra.
\end{equation}

\subsection{Unequal-time commutator}

Next we calculate the unequal-time commutator $[x(t_1), x(t_2)]$ since this object determines the linear response of the system to external perturbations~\cite{Kubo:1957mj}
\begin{equation}\label{exact_com}
    \begin{split}
    [x(t_1), x(t_2)] &= \frac{1}{2 f_+ \sqrt{1 - 4 \l \o^2}}
    ([a_+ , a_+^\da] e^{- i f_+ (t_1 - t_2)} + [a_+^\da , a_+] e^{i f_+ (t_1 - t_2)}) \\
    & \quad + \frac{1}{2 f_- \sqrt{1 - 4 \l \o^2}}
    ([a_-^\da , a_-] e^{- i f_- (t_1 - t_2)} 
    + [a_- , a_-^\da] e^{i f_- (t_1 - t_2)})  \\
    &= \frac{-i}{f_+ \sqrt{1 - 4 \l \o^2}} 
    \sin [f_+ (t_1 - t_2)]
    + \frac{i}{f_- \sqrt{1 - 4 \l \o^2}} 
    \sin [f_- (t_1 - t_2)].
    \end{split}
\end{equation}

\subsection{Behavior as $\l \to 0$}

We consider the behavior of this model as $\lambda \rightarrow 0$. In the limit, $\lambda\rightarrow 0$, $f_{+}\sim\omega$, $f_{-}\sim \infty$. The behavior $f_{-}\sim \infty$ should not be interpreted literally. It only reflects some fast modes at the energy scale of $f_-$ which are beyond the model's energy scale. The decoupling of these fast modes follows the same reasons discussed in~\cite{Wu:2026xgd}: classically their frequencies are very large and their contributions to any macroscopic observable are ignorable; quantum mechanically, their excitation energies lie far beyond the low energy cutoff. We therefore restrict the perturbative expansion to the slow mode $f_+$ in what follows.

We ignore the fast mode and measure the characteristic frequency $f_+$ with accuracy up to $\mc{O}(\l^2)$, then
\begin{equation}
    f_+ = \o(1 + \tfrac{1}{2} \l \o^2 + \tfrac{7}{8} \l^2 \o^4) + \mc{O}(\l^3).
\end{equation}
Thus the energy spectrum becomes
\begin{equation}\label{H_expan}
    H | n_+ \ra = \o (1 + \tfrac{1}{2} \l \o^2 + \tfrac{7}{8} \l^2 \o^4) (n_+ + \tfrac{1}{2}) 
    | n_+ \ra + \mc{O}(\l^3).
\end{equation}
The fast mode in the unequal-time commutator is also smeared out over a macroscopic time interval. The $[x(t_1), x(t_2)]$ up to $\mc{O}(\l)$ reads
\begin{equation}\label{com_expan}
    [x(t_1), x(t_2)] = \tfrac{-i}{\o} 
    (1 + \tfrac{3}{2} \l \o^2) 
    \sin [\o (t_1 - t_2)] 
    - \tfrac{i}{2} \l \o^2 (t_1 - t_2) 
    \cos[\o (t_1 - t_2)]
    + \mc{O}(\l^2).
\end{equation}
In the following discussion, the expansion of the exact low-energy sector will serve as a rigorous benchmark against which the perturbative results should be tested.


\section{Perturbative quantization with covariant phase space formalism}\label{Sec4}

In this section we quantize this model using the perturbation method in the framework of CPS\footnote{The equivalence between the CPS formalism and the standard Ostrogradsky's construction is established in 
App.~\ref{app:equivalence}.}. By constructing the symplectic form directly on the low-energy sector of the solution space, the ghost mode is automatically decoupled. And the perturbative results agree with the expansion of the low-energy branch of the exact theory.

\subsection{Symplectic form and Hamiltonian}

The variation of the Lagrangian form $L = \mc{L} \d t$ is  
\begin{equation}
    \de L = -(\l \ddddot{x} + \ddot{x} + \o^2 x) \de x \d t
    + \d \Theta,
\end{equation}
where the symplectic potential $\Theta$ is
\begin{equation}
    \Theta = (\dot{x} + \l \dddot{x} )\de x 
    - \l \ddot{x} \de \dot{x}.
\end{equation}
We then calculate the symplectic form
\begin{equation}\label{SF_x}
    \O = \de \Theta = (\de\dot{x} + \l \de \dddot{x} )
    \w \de x 
    - \l \de\ddot{x} \w \de \dot{x}.
\end{equation}
And the Hamiltonian is~\cite{Iyer:1994ys,Harlow:2019yfa}
\begin{equation}\label{Ham_x}
    H \equiv J_\xi = X_\xi \cdot \Theta - \xi \cdot L
    = \frac{1}{2} \o^2 x^2 + \frac{1}{2} \dot{x}^2 
    - \frac{1}{2} \l \ddot{x}^2 
    + \l \dot{x} \dddot{x},
\end{equation}
where $\xi \equiv \frac{\del}{\del t}$ stands for time evolution, $X_\xi$ is defined as $X_\xi \equiv \mc{L}_\xi x \frac{\de}{\de x} + \mc{L}_\xi y \frac{\de}{\de y}$, and $\cdot$ denotes
inserting a vector into the first argument of a differential form.

\subsection{Perturbative solution}

Suppose the solution to EOM can be written as the following perturbative expansion
\begin{equation}\label{pert_seri}
    x(t) = x_0(t) + \l x_1(t) + \l^2 x_2(t) + \cdots,
\end{equation}
where
\begin{equation}
    x_0 (t) = \tfrac{1}{\sqrt{2 \o}} 
    (a e^{-i \o t} + a^\da e^{i \o t})
\end{equation}
is the solution to the unperturbed EOM (\ref{EOM}), and $x_1(t)$, $x_2(t)$ are corrections remain to be solved in the perturbation theory. We substitute this expansion into the EOM. To first order in $\l$ we obtain
\begin{equation}
    \ddot{x}_1(t) + \o^2 x_1 (t) = -\ddddot{x}_0(t) 
    = -\o^4 x_0(t).
\end{equation}
One specific solution is
\begin{equation}
    x_1 (t) = \tfrac{3}{4} \o^2 x_0 (t) 
    + \tfrac{1}{2} \o^2 t \dot{x}_0(t).
\end{equation}
Thus the solution valid to first order in $\l$ is
\begin{equation}\label{1st_per_sol}
    x (t) = (1 + \tfrac{3}{4} \l \o^2) x_0(t) 
    + \tfrac{1}{2} \l \o^2 t \dot{x}_0(t) + \mc{O}(\l^2).
\end{equation}
We then expand the EOM to second order
\begin{equation}
    \ddot{x}_2(t) + \o^2 x_2(t) = -  \ddddot{x}_1 (t)
    = - \tfrac{11}{4} \o^6 x_0(t) - \tfrac{1}{2} \o^6 t \dot{x}_0(t),
\end{equation}
and one specific solution is
\begin{equation}
    x_2(t) = \tfrac{61}{32} \o^4 x_0(t) 
    + \tfrac{5}{4} \o^4 t \dot{x}_0(t)
    - \tfrac{1}{8} \o^6 t^2 x_0 (t).
\end{equation}
Therefore we obtain the perturbative solution valid to second order\footnote{This solution is only valid up to a finite time interval $|t| \ll \frac{1}{\l \o^3}$. The specific solutions $x_1(t)$ and $x_2(t)$ are determined such that operators $a$ and $a^\da$ satisfy the standard creation-annihilation algebra upon quantization.}
\begin{equation}\label{2nd_pert_sol}
    x(t) = (1 + \tfrac{3}{4} \l \o^2 
    + \tfrac{61}{32} \l^2 \o^4)x_0(t) 
    +  \tfrac{1}{2} \l \o^2 t \dot{x}_0(t)
    + \tfrac{5}{4} \l^2 \o^4 t \dot{x}_0(t)
    - \tfrac{1}{8} \l^2 \o^6 t^2 x_0 (t) 
    + \mc{O}(\l^3).
\end{equation}

\subsection{Quantization}\label{Pert_Quan}

We substitute the perturbative solution Eq.~(\ref{2nd_pert_sol}) into the symplectic form and obtain (see App.~\ref{app: SF} for detailed computations)
\begin{equation}\label{SF_a}
    \O = -i \de a \w \de a^\da + \mc{O}(\l^3).
\end{equation}
To quantize the system, we promote $a$ and $a^\da$ to operators, and impose the following commutation relation
\begin{equation}\label{CPS_quant}
    [a, a^\da] = i \O^{-1} (\de a, \de a^\da) = 1.
\end{equation}
We then substitute the perturbative solution into the Hamiltonian, and obtain (detailed in App.~\ref{app: H})
\begin{equation}\label{Ham_a}
    H = \o (1 + \tfrac{1}{2} \l \o^2 + \tfrac{7}{8} \l^2 \o^4) (a^\da a + \tfrac{1}{2}) + \mc{O}(\l^3).
\end{equation}
We introduce number operator $N = a^\da a$ and label its eigenstate by $| n \ra$. Thus the energy spectrum is
\begin{equation}
    H | n \ra = \o (1 + \tfrac{1}{2} \l \o^2 + \tfrac{7}{8} \l^2 \o^4) (n + \tfrac{1}{2}) | n \ra
    + \mc{O}(\l^3).
\end{equation}
This result matches the expansion of the exact energy spectrum Eq.~(\ref{H_expan}).

\subsection{Unequal-time commutator}

Based on the perturbative solution Eq.~(\ref{1st_per_sol}) and the commutation relation Eq.~(\ref{CPS_quant}), we calculate the unequal-time commutator up to $\mc{O}(\l)$
\begin{equation}
\begin{split}
    [x(t_1), x(t_2)] 
    &= (1 + \tfrac{3}{4} \l \o^2)^2 [x_0 (t_1), x_0(t_2)] \\
    & \quad + \tfrac{1}{2} \l \o^2 t_2
    [x_0 (t_1) , \dot{x}_0 (t_2)]
    +  \tfrac{1}{2} \l \o^2 t_1
    [\dot{x}_0 (t_1) , x_0 (t_2) ] + \mc{O}(\l^2) \\
    &= \tfrac{-i}{\o}  (1 + \tfrac{3}{2} \l \o^2) \sin[\o (t_1 - t_2)]
    - \tfrac{i}{2} \l \o^2 (t_1 - t_2) \cos[\o (t_1 - t_2)]
    +\mc{O}(\l^2).
\end{split}
\end{equation}
This result is the same as the expansion of the slow sector of the exact unequal-time commutator Eq.~(\ref{com_expan}).


\section{Conclusion and discussion}\label{Sec5}

In this paper, we have applied the covariant phase space formalism to the perturbative quantization of the higher-derivative oscillator. We treated the higher-derivative term as a perturbation and constructed the symplectic form directly on the low-energy solution space. The quantization procedure yielded a positive-definite Hamiltonian whose spectrum matches the expansion of the low-energy sector of the exact theory. The Ostrogradsky ghost, which obstructs the standard canonical quantization, is automatically absent from the perturbative construction.

This work demonstrates that the CPS formalism, already shown to be effective for perturbations that alter the constraint structure~\cite{Wu:2026xgd}, also provides a clean and systematic treatment of genuine higher-derivative perturbations. The decoupling of the ghost mode is not achieved by imposing constraints or by manual truncation, but emerges naturally from the geometry of the solution space. We hope that the CPS approach may offer a promising route toward the quantization of more complex higher-derivative theories.


\section*{Acknowledgments}

We thank Bin Chen, Pratik Rath, Guanhao Sun, Cong Zhang and Hongbao Zhang for helpful discussions.
This work is supported by the National Natural Science Foundation of China (NSFC) Project No.12447101 and No.12575079.


\appendix

\section{Equivalence between the CPS and the Ostrogradsky's construction}\label{app:equivalence}

In this appendix we establish the equivalence between the covariant phase space formalism and the Ostrogradsky's construction for general non-degenerate higher-derivative theories. 

\subsection{Ostrogradsky's construction}

Consider a Lagrangian that depends on time derivatives of a single coordinate $q(t)$ up to order $N$,
\begin{equation}\label{gen_L}
    \mathcal{L} = \mathcal{L}(q, \dot{q}, \ddot{q}, \dots, q^{(N)}), \qquad 
    \frac{\partial^2 \mathcal{L}}{\partial q^{(N)2}} \neq 0,
\end{equation}
where the non-degeneracy condition guarantees that the Euler‑Lagrange equation is of order $2N$.

The Ostrogradsky's construction introduces the canonical coordinates\cite{Woodard:2015zca}
\begin{align}
    Q_i &\equiv q^{(i-1)}, \label{Ost_Q} \\
    P_i &\equiv \sum_{j=i}^{N} \left( -\frac{\d}{\d t} \right)^{\!j-i}
    \frac{\partial \mathcal{L}}{\partial q^{(j)}},  \label{Ost_P}
\end{align}
where $i = 1,\dots,N$. The non-degeneracy condition guarantees that $P_N = \partial \mathcal{L} / \partial q^{(N)}$ can be inverted to solve for the highest derivative $q^{(N)}$ as a function of $(Q_1,\dots,Q_N, P_N)$. This is the crucial step that allows the Legendre transform to be completed.

The Ostrogradsky Hamiltonian is defined by the  Legendre transform
\begin{equation}\label{H_os}
    H_{\mathrm{Os}}(Q_1,\dots,Q_N,P_1,\dots,P_N) 
    = \sum_{i=1}^{N} P_i \dot{Q}_i - \mathcal{L},
\end{equation}
where the velocities $\dot{Q}_i$ are understood to be expressed in terms of the canonical variables.

\subsection{Isomorphism of phase spaces }

We now apply the CPS formalism directly to the higher‑derivative Lagrangian. 
The Euler‑Lagrange equation derived from the non‑degenerate Lagrangian Eq.~(\ref{gen_L}) are of order $2N$. Their solutions are therefore uniquely determined by the initial data
\begin{equation}
    \bigl( q, \dot{q}, \ddot{q}, \dots, q^{(2N-1)} \bigr) \big|_{t=t_0}.
\end{equation}
These $2N$ real numbers provide a coordinate on the CPS phase space $\mathcal{P}$. Through the definitions (\ref{Ost_Q}) and (\ref{Ost_P})
they also determine the initial values of the $2N$ Ostrogradsky's canonical variables $(Q_i, P_i)$. This establishes a diffeomorphism between  the CPS phase space $\mathcal{P}$ and the Ostrogradsky's phase space 
$\mathcal{P}_{\mathrm{Os}}$.

\subsection{Equivalence of symplectic structures}

The variation of the Lagrangian form $L = \mathcal{L}\d t$ is
\begin{equation}
    \delta L = E  \delta q \d t + \d\Theta,
\end{equation}
where $E$ is the equation of motion, and the symplectic potential $\Theta$ is a $(d-1)$-form on the solution space. A standard computation yields
\begin{equation}\label{Theta_general}
    \Theta = \sum_{i=1}^{N} \left[ \sum_{j=i}^{N} 
    \left( -\frac{\d}{\d t} \right)^{\!j-i}
    \frac{\partial \mathcal{L}}{\partial q^{(j)}} \right] 
    \delta q^{(i-1)}.
\end{equation}
Comparing with Eqs.~(\ref{Ost_Q}) and (\ref{Ost_P}) one immediately recognizes that
\begin{equation}\label{Theta_PQ}
    \Theta = \sum_{i=1}^{N} P_i \, \delta Q_i.
\end{equation}
The symplectic form is obtained by taking the exterior derivative:
\begin{equation}
    \Omega = \delta\Theta = \sum_{i=1}^{N} \delta P_i \wedge \delta Q_i,
\end{equation}
which is identical to the canonical symplectic structure defined by Poisson brackets on the Ostrogradsky's phase space.

\subsection{Equivalence of Hamiltonians}

The CPS Hamiltonian is the Noether charge associated with time translation $\xi = \partial_t$\cite{Iyer:1994ys,Harlow:2019yfa}
\begin{equation}
    H_{\mathrm{CPS}} = X_\xi \cdot \Theta - \xi \cdot L,
\end{equation}
where $ X_\xi \equiv \mc{L}_\xi x \frac{\de}{\de x} + \mc{L}_\xi y \frac{\de}{\de y}$ is the vector field on the solution space that generates time evolution.
Using the expression (\ref{Theta_PQ}) for $\Theta$,
we obtain
\begin{equation}
    H_{\mathrm{CPS}} = 
    \sum_{i=1}^{N} P_i \dot{Q}_i - \mathcal{L}
    = H_{\mathrm{Os}}.
\end{equation}
Therefore the two on-shell Hamiltonians coincide.

\subsection{Summary}

The covariant phase space formalism, when applied to a 
non‑degenerate higher‑derivative Lagrangian, produces the canonical structure that is identical to the Ostrogradsky's construction. This guarantees that the CPS quantization developed in the main text captures the same low-energy physics as the standard Ostrogradsky quantization.

\section{Calculation of the symplectic form}\label{app: SF}

In this appendix we calculate the symplectic form Eq.~(\ref{SF_a}) based on the perturbative solution Eq.~(\ref{2nd_pert_sol}).

The symplectic form for the oscillator is given by Eq.~(\ref{SF_x}). Using the perturbative series Eq.~(\ref{pert_seri}), we write $\Omega = \Omega^{(0)} + \lambda \O^{(1)} + \l^2 \O^{(2)}$, where 
$\Omega^{(0)} = \de \dot{x}_0 \w \de x_0$ is the symplectic form of the unperturbed theory, and 
\begin{equation}
    \Omega^{(1)} = \delta \dot{x}_0 \wedge \delta x_1 + \delta \dot{x}_1 \wedge \delta x_0
    - \delta \ddot{x}_0 \wedge \delta \dot{x}_0 + \delta \dddot{x}_0 \wedge \delta x_0,
\end{equation}
and
\begin{align}
    \Omega^{(2)} ={}& \delta \dot{x}_0 \wedge \delta x_2 + \delta \dot{x}_1 \wedge \delta x_1 + \delta \dot{x}_2 \wedge \delta x_0 \notag \\
    &- \delta \ddot{x}_0 \wedge \delta \dot{x}_1 - \delta \ddot{x}_1 \wedge \delta \dot{x}_0 \notag \\
    &+ \delta \dddot{x}_0 \wedge \delta x_1 + \delta \dddot{x}_1 \wedge \delta x_0
\end{align}
are corrections to $\Omega^{(0)}$.

Using $x_1 = \frac{3}{4}\omega^2 x_0 + \frac{1}{2}\omega^2 t \dot{x}_0$ and $\ddot{x}_0 = -\omega^2 x_0$, we calculate
\begin{equation}
    \Omega^{(1)} = 
    \tfrac{3}{4}\o^2 \de \dot{x}_0 \w \de x_0
    + \tfrac{5}{4} \o^2 \de \dot{x}_0 \w \de x_0
    + \o^2 \de x_0 \w \de \dot{x}_0 
    - \o^2 \de \dot{x}_0 \w \de x_0
    = 0 .
\end{equation}
Substituting the explicit expressions for $x_1$ and $x_2$ and using identities
\begin{align}
    \ddot{x}_0 &= -\omega^2 x_0,  \\
    \dot{x}_1 &= \tfrac{5}{4} \o^2 \dot{x}_0 
    - \tfrac{1}{2} \o^4 t x_0  \\
    \ddot{x}_1 &= - \tfrac{7}{4}\omega^4 x_0 - \tfrac{1}{2}\omega^4 t \dot{x}_0, \\
    \dddot{x}_1 &= -\tfrac{9}{4}\omega^4 \dot{x}_0 
    + \tfrac{1}{2}\omega^6 t x_0,
\end{align}
we obtain
\begin{equation}
\begin{split}
    \Omega^{(2)} &= \de \dot{x}_0 \w 
    (\tfrac{61}{32} \o^4 \de x_0 - \tfrac{1}{8} \o^6 t^2 \de x_0)
    +(\tfrac{5}{4} \o^2 \de \dot{x}_0 - \tfrac{1}{2} \o^4 t \de x_0) \w
    (\tfrac{3}{4} \o^2 \de x_0 + \tfrac{1}{2} \o^2 t \de \dot{x}_0)  \\
    & \quad + [(\tfrac{61}{32} \o^4 + \tfrac{5}{4} \o^4) \de \dot{x}_0 - \tfrac{1}{8} \o^6 t^2 \de \dot{x}_0] 
    \w \de x_0  \\
    & \quad + \o^2 \de x_0 \w \tfrac{5}{4} \o^2 \de \dot{x}_0 
    - ( - \tfrac{7}{4} \o^4 \de x_0) \w \de \dot{x}_0 \\
    & \quad - \o^2 \de \dot{x}_0 \w \tfrac{3}{4} \o^2 \de x_0 + (-\tfrac{9}{4} \o^4 \de \dot{x}_0) \w \de x_0 \\
    &= [- 2 * \tfrac{61}{32} \o^4 + \tfrac{1}{8} \o^6 t^2 - \tfrac{15}{16} \o^4 -\tfrac{1}{4} \o^6 t^2 - \tfrac{5}{4} \o^4    \\
    &\quad + \tfrac{1}{8} \o^6 t^2 + \tfrac{5}{4} \o^4 + \tfrac{7}{4} \o^4 + \tfrac{3}{4} \o^4 + \tfrac{9}{4} \o^4] \de x_0 \w \de \dot{x}_0 \\
    &= 0.
\end{split}
\end{equation}
Therefore,
\begin{equation}
    \Omega = \delta \dot{x}_0 \wedge \delta x_0 + \mathcal{O}(\lambda^3) .
\end{equation}
Finally, because
\begin{equation}
    x_0 = \tfrac{1}{\sqrt{2\omega}} (a e^{-i\omega t} + a^\dagger e^{i\omega t}), \qquad
    \dot{x}_0 = -i\sqrt{\tfrac{\omega}{2}} (a e^{-i\omega t} - a^\dagger e^{i\omega t}),
\end{equation}
we obtain
\begin{equation}
    \Omega = -i\,\delta a \wedge \delta a^\dagger + \mathcal{O}(\lambda^3).
\end{equation}

\section{Calculation of the Hamiltonian}\label{app: H}

In this appendix we calculate the Hamiltonian Eq.~(\ref{Ham_a}).

Since the Hamiltonian does not evolve with time, given the second-order perturbative solution Eq.~(\ref{2nd_pert_sol}), we evaluate
\begin{equation}
    x(0) = (1 + \tfrac{3}{4} \l \o^2 + \tfrac{61}{32} \l^2 \o^4) x_0(0) + \mc{O}(\l^3),
\end{equation}
\begin{equation}
    \dot{x}(0) = (1 + \tfrac{5}{4} \l \o^2 + \tfrac{101}{32} \l^2 \o^4) \dot{x}_0(0) + \mc{O}(\l^3),
\end{equation}
\begin{equation}
    \ddot{x}(0) = -\o^2 (1 + \tfrac{7}{4} \l \o^2) x_0(0) + \mc{O}(\l^2),
\end{equation}
and
\begin{equation}
    \dddot{x}(0) = - \o^2 (1 + \tfrac{9}{4} \l \o^2) \dot{x}_0(0) + \mc{O}(\l^2).
\end{equation}
Thus
\begin{equation}
\begin{split}
    H &= \tfrac{\o^2}{2} x^2 + \tfrac{1}{2} \dot{x}^2
    - \tfrac{\l}{2} \ddot{x}^2 + \l \dot{x} \dddot{x} \\
    & = \tfrac{\o^2}{2} (1 + \tfrac{3}{2} \l \o^2 + \tfrac{35}{8} \l^2 \o^4) x_0^2 
    +\tfrac{1}{2} (1 + \tfrac{5}{2} \l \o^2 + \tfrac{63}{8}\l^2 \o^4) \dot{x}_0^2 \\
    &\quad - \tfrac{\l}{2} \o^4 (1 + \tfrac{7}{2} \l \o^2) x_0^2 - \l \o^2 (1 + \tfrac{7}{2} \l \o^2)\dot{x}_0^2 + \mc{O}(\l^3) \\
    &= \tfrac{\o^2}{2} [((1 + \tfrac{3}{2} \l \o^2 + \tfrac{35}{8} \l^2 \o^4)) - \l \o^2 (1 + \tfrac{7}{2} \l \o^2) ] x_0^2 \\
    &\quad + \tfrac{1}{2} [(1 + \tfrac{5}{2} \l \o^2 + \tfrac{63}{8}\l^2 \o^4) - 2\l \o^2 (1 + \tfrac{7}{2} \l \o^2)] \dot{x}_0^2+ \mc{O}(\l^3)  \\
    &= \tfrac{\o^2}{2} (1 + \tfrac{1}{2} \l \o^2 + \tfrac{7}{8} \l^2 \o^4) (x_0^2 + \tfrac{1}{\o^2} \dot{x}_0^2) + \mc{O}(\l^3) \\
    &= \o (1 + \tfrac{1}{2} \l \o^2 + \tfrac{7}{8} \l^2 \o^4) (a^\da a + \tfrac{1}{2}) + \mc{O}(\l^3).
\end{split}
\end{equation}


\addcontentsline{toc}{section}{References}

\bibliographystyle{JHEP}
\bibliography{refs}

\end{document}